\documentstyle[lprocl,epsfig,11pt]{article}
\textheight
8.5in  \parskip 7pt \openup1\jot 
\def\st{\scriptstyle}

\def\be{\begin{equation}}
\def\ee{\end{equation}}
\def\bea{\begin{eqnarray}}
\def\eea{\end{eqnarray}}

\newskip\humongous \humongous=0pt plus 1000pt minus 1000pt
\def\caja{\mathsurround=0pt}
\def\eqalign#1{\,\vcenter{\openup1\jot \caja
        \ialign{\strut \hfil$\displaystyle{##}$&$
        \displaystyle{{}##}$\hfil\crcr#1\crcr}}\,}
\newif\ifdtup

\def\eqright #1\cr{\noalign{\hfill$\displaystyle{{}#1}$}}
\def\eqleft #1\cr{\noalign{\noindent$\displaystyle{{}#1}$\hfill}}

\def\oldreffmt#1{\rlap{[#1]} \hbox to 2\parindent{}}

\def\figfmt#1{\rlap{Figure {#1}} \hbox to 1in{}}

\def\VEV#1{\left\langle #1\right\rangle}

\def\begineq #1\endeq{$$ \refstepcounter{equation}\eqalign{#1}\eqno
	(\theequation) $$}
\def\contlimit{\,{\hbox{$\longrightarrow$}\kern-1.8em\lower1ex
\hbox{${\scriptstyle (a\rightarrow0)}$}}\,}
\def\centeron#1#2{{\setbox0=\hbox{#1}\setbox1=\hbox{#2}\ifdim
\wd1>\wd0\kern.5\wd1\kern-.5\wd0\fi
\copy0\kern-.5\wd0\kern-.5\wd1\copy1\ifdim\wd0>\wd1
\kern.5\wd0\kern-.5\wd1\fi}}
\def\centerover#1#2{\centeron{#1}{\setbox0=\hbox{#1}\setbox
1=\hbox{#2}\raise\ht0\hbox{\raise\dp1\hbox{\copy1}}}}
\def\centerunder#1#2{\centeron{#1}{\setbox0=\hbox{#1}\setbox
1=\hbox{#2}\lower\dp0\hbox{\lower\ht1\hbox{\copy1}}}}
\def\lsim{\;\centeron{\raise.35ex\hbox{$<$}}{\lower.65ex\hbox
{$\sim$}}\;}
\def\gsim{\;\centeron{\raise.35ex\hbox{$>$}}{\lower.65ex\hbox
{$\sim$}}\;}
\def\st#1{\centeron{$#1$}{$/$}}

\def\super#1{\ifmmode \hbox{\textsuper{#1}}\else\textsuper{#1}\fi}
\def\textsuper#1{\newcount\holdspacefactor\holdspacefactor=\spacefactor
$^{#1}$\spacefactor=\holdspacefactor}

\def\getcite#1,{\advance\citenumber by1
\ifnum\citenumber=1
\ref{#1}\let\next=\getcite\else\ifx#1@\let\next=\relax
\else ,\ref{#1}\let\next=\getcite\fi\fi\next}

\def\upon #1/#2 {{\textstyle{#1\over #2}}}
\relax

\def\til#1{\centeron{\hbox{$#1$}}{\lower 2ex\hbox{$\char'176$}}}
\def\tild#1{\centeron{\hbox{$\,#1$}}{\lower 2.5ex\hbox{$\char'176$}}}
\def\sumtil{\centeron{\hbox{$\displaystyle\sum$}}{\lower
-1.5ex\hbox{$\widetilde{\phantom{xx}}$}}}

\def\pom{{\rm P\kern -0.53em\llap I\,}}
\def\spom{{\rm P\kern -0.36em\llap \small I\,}}
\def\sspom{{\rm P\kern -0.33em\llap \footnotesize I\,}}

\begin{document} 

\begin{titlepage} 

\rightline{\vbox{\halign{&#\hfil\cr
&ANL-HEP-CP-97-68 \cr
&\today\cr}}} 
\vspace{1.25in} 

\begin{center}
{\bf DEEP-INELASTIC DIFFRACTION AND THE POMERON AS A SINGLE GLUON 
}\footnote{Work 
supported by the U.S.
Department of Energy, Division of High Energy Physics, \newline Contracts
W-31-109-ENG-38 and DEFG05-86-ER-40272} 
\medskip

Alan. R. White\footnote{arw@hep.anl.gov }
\end{center}
\vskip 0.6cm

\centerline{High Energy Physics Division}
\centerline{Argonne National Laboratory}
\centerline{9700 South Cass, Il 60439, USA.}
\vspace{0.5cm}

\begin{abstract} 

Deep-inelastic diffractive scaling provides fundamental insight into the QCD
pomeron. It is argued that single gluon domination of the structure
function, together with the well-known Regge pole property, determines that
the pomeron carries color-charge parity $ C_c = -1~$ and, at short
distances, is in a super-critical phase of Reggeon Field Theory. The main
purpose of the talk is to describe the relationship of the super-critical
pomeron to QCD. 

\end{abstract} 

\vspace{2in}
\begin{center}

Presented at the International Conference (VIIth Blois Workshop) On Elastic
and Diffractive Scattering - Recent Advances in Hadron Physics.
\newline  Seoul, Korea. June 10-14, 1997.

\end{center}

\end{titlepage}

\section{Introduction}

Understanding the pomeron in QCD is equivalent to solving the theory at 
high-energy. In this talk I want to focus on two, apparently very different,
experimental properties of the pomeron that I believe provide important 
insight into the problem. 

(i) 35 years of experiment/phenomenology/theory 
have shown that the pomeron is, 
\newline \indent approximately, a Regge pole.  

(ii) Recent DIS diffractive scaling violation 
results~\cite{h1} show that (at $Q^2 \sim 5 ~GeV^2$) the 
\newline \indent pomeron is, approximately, a single
gluon. 

\noindent It has been known for a long time that (i) is very difficult to
reconcile with perturbative QCD and the 2 gluon BFKL pomeron. (ii) is a 
relatively new 
result that, as I discuss further below, is similarly in conflict 
with leading-twist perturbative QCD (and even, at first sight, with gauge
invariance). In this talk I will argue that (i) and (ii) are closely related
and reflect a subtle mixture of perturbative and non-perturbative physics in
the Regge limit. We will see that both (i) and (ii) are satisfied if the
pomeron carries color-charge parity $ C_c = -1~$ and, at short distances, is
in a super-critical phase of Reggeon Field Theory. The focus of the talk
will be on the relationship of the super-critical pomeron to QCD. 

It was more than sixteen years ago that I first suggested~\cite{arw} the pomeron
could appear, in a super-critical phase, as a single (reggeized) gluon in a
soft gluon background. Subsequently~\cite{arw1,arw2}, I laid out what I hoped
might be the basis for a full dynamical understanding of the QCD pomeron. 
Although my arguments were very incomplete, they implied a fundamentally
different picture of the pomeron to what might be called the conventional
BFKL perturbative picture. 
It was clear that in my picture the nature of the pomeron is
intimately related to confinement and chiral symmetry 
breaking. The scaling violation results~\cite{h1} from H1
have encouraged me to return to this work and attempt to put it on 
firmer ground. I briefly outline below the central elements of a new 
paper in preparation. Although the global picture 
presented in~\cite{arw1} re-emerges, the details are 
different in important ways. 

\section{Outline}

\noindent I begin, in Section 3, by reviewing the DIS diffractive results
and the H1 analysis of the scaling violations. 
In Section 4, I go on to discuss why the large logarithmic violations can not
be reproduced by conventional leading-twist perturbative QCD calculations. 
I then list, very briefly, in Section 5 the elements of multi-Regge
theory~\cite{arw2} that provide the underlying basis for my analysis. 
The main body of of the talk, Sections 6, 7 and 8, outlines how multi-regge
theory can be used to simultaneously derive both 
the dynamical pomeron and hadron reggeons via super-critical RFT. Finally, 
in Section 9, I briefly discuss how the single gluon approximation appears 
in DIS diffraction. 

\newpage

The essential points of the talk are the following.

\begin{itemize}

\item {Existing Regge limit QCD calculations~\cite{ln}
, together with new calculations of 
further ``helicity-flip'' reggeon vertices allow 
multi-Regge theory to be used to 
construct the asymptotic behavior of very 
complicated scattering processes in terms of $J$-plane reggeon diagrams 
($k_{\perp}$ integrals with reggeon propagators $[J - 1 + \sum \Delta 
(k_{\perp i}^2) ]^{-1}$ and interactions). This description is unitary (and 
complete) in the small $k_{\perp}$ region when gluons are massive. 
}

\item {In such processes bound states and their
scattering amplitudes can be simultaneously studied and the infra-red 
(massless gluon) limit taken. } 
\vspace{0.15in}

\item {In helicity-flip multi-Regge kinematics, perturbative massless 
quark contributions to vertices contain the infra-red triangle anomaly and 
produce the chirality violation normally associated with non-perturbative
instanton interactions. }

\item {An effective multi-regge reggeon 
theory of (massless) quarks and gluons that implicitly includes 
effects of instanton interactions can be constructed. }
\vspace{0.15in}

\item {In the infra-red limit giving SU(2) gauge symmetry 
(recall that an instanton interaction is associated with an SU(2)
subgroup), infra-red divergent multi-regge diagrams appear which contain
``hadron'' reggeon states scattering via ``pomeron'' exchange. }

\item {The pomeron appears as a Regge pole which is, in first 
approximation, a reggeized gluon in a ``regggeon condensate''. All the
features~\cite{arw2} of super-critical RFT are present. } 

\item {Hadron reggeons have a 
confinement and chiral symmetry breaking spectrum and (although details 
remain to be worked out it appears that) chiral symmetry 
breaking is essential for the self-consistency of the pomeron.
}

\item {When there is a $k_{\perp}$ cut-off, ``complimentarity'' implies the
full SU(3) gauge symmetry can be smoothly restored. Critical behavior of the
Pomeron is involved. } 

\item {In DIS diffraction the  reggeized gluon 
appears as a single gluon.}
 
\end{itemize}

\newpage

\section{DIS Diffractive Scaling Violations } 

\noindent The definition of the diffractive structure function 
$F^D_2(x_{\spom},\beta,Q^2)$ is illustrated in Fig.~3.1
\parbox{3in} 
{ \openup\jot We use the usual kinematic variables {\small
$$
W^2 = (P+Q)^2~~ ~
\beta = {Q^2 \over Q^2 + M_X^2}~~~
x_{\spom} = {Q^2 + M_X^2 \over Q^2 + W^2} 
$$
} Fig.~3.2 shows the H1 results~\cite{h1}
for the $Q^2$ and $\beta$ dependence of 
$F^D_2(x_{\spom},\beta,Q^2)$ at small-$x_{\spom}$. 
The presence of positive $\ln [Q^2]$ scaling violations 
over a large range of $\beta$ is clear.
}
\parbox{0.2in}{$~$}
\parbox{1.6in}{
\begin{center}
\leavevmode
\epsfxsize=1.4in
\epsffile{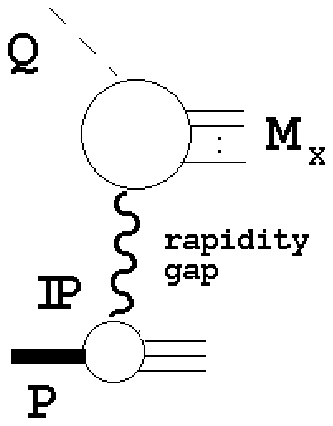} 
\end{center}
 }
\parbox{1.2in}{$ \to F^D_2(x_{\spom},\beta,Q^2) $}
\newline \parbox{3.9in}{ 
\epsfig{file=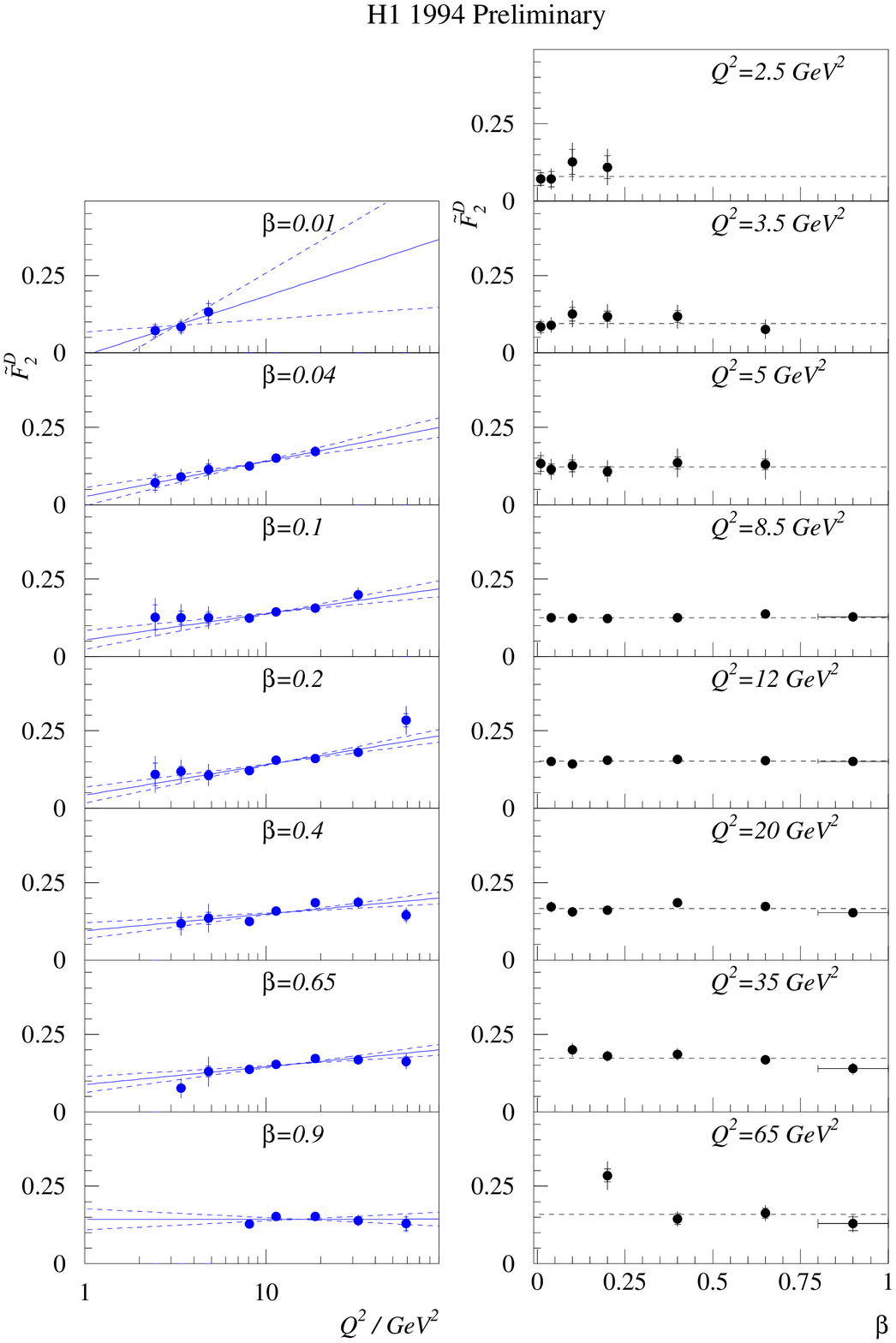,bbllx=0pt,bblly=150pt,bburx=600pt,bbury=870pt,%
rwidth=0.1in,rheight=5.2in,width=3.6in} 
\centerline{Fig.~3.2} 
}\parbox{2.1in}{ \openup\jot \centerline{Fig.~3.1}
$~$
\newline 
By fitting to DGLAP evolution, H1 have extracted~\cite{h1}  
the low $Q^2$ pomeron structure function shown in Fig.~3.3. To reproduce
(at medium $\beta$) the 
logarithmic rise at large $Q^2$, {\it a 
single gluon } must carry nearly all the pomeron momentum. 
\newline $~$

\epsfig{file=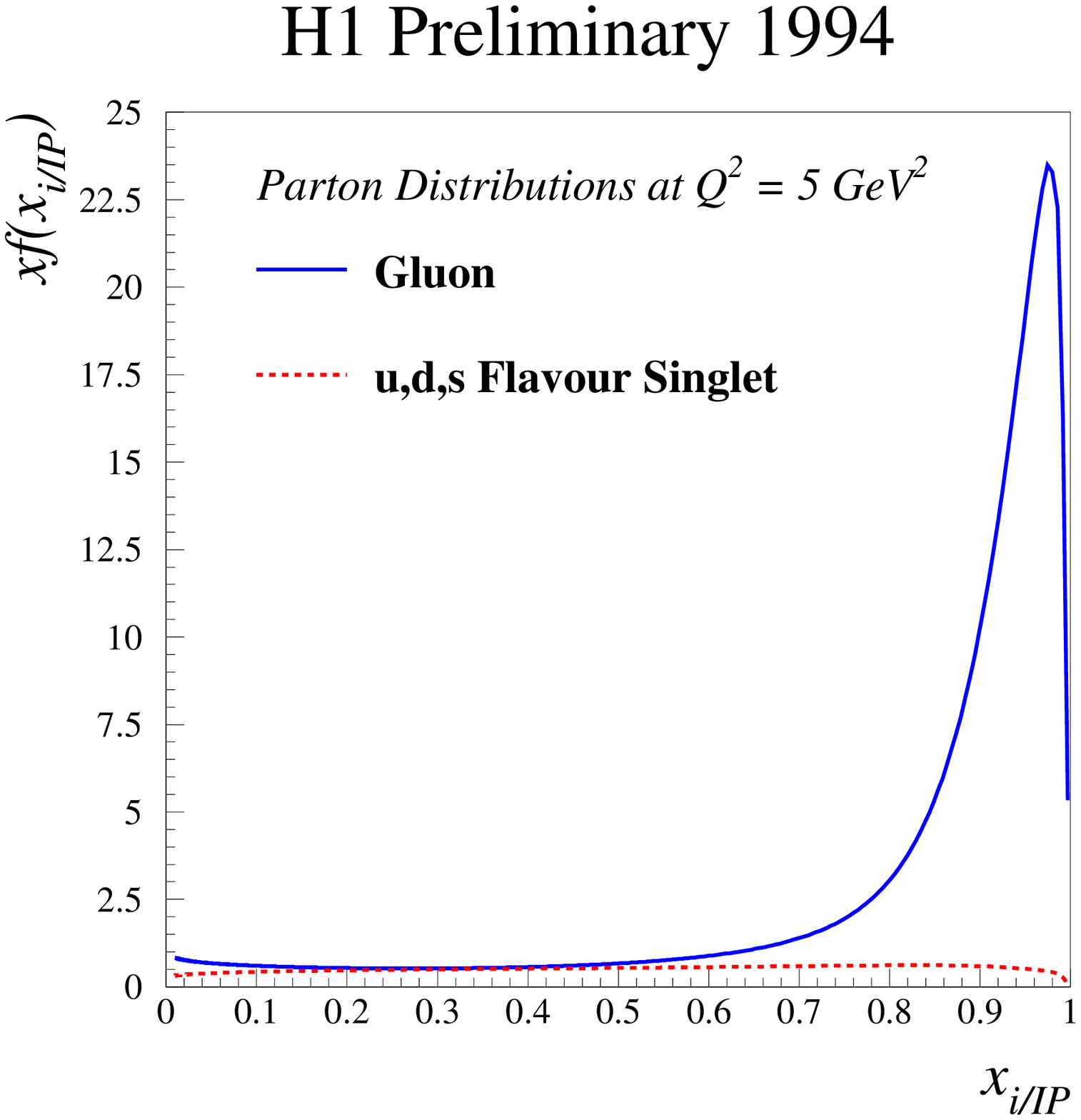,bbllx=40pt,bblly=0pt,bburx=590pt,bbury=550pt,%
width=2.4in,rwidth=0.2in}

\centerline{Fig.~3.3} 
}

\noindent As we discuss next, for perturbative two 
gluon exchange, a single gluon can not carry nearly all the momentum.
This is closely related to the 
scale-invariance property of the BFKL pomeron which produces a 
fixed singularity in the $J$-plane rather than a Regge pole.

\section{Perturbative Scaling Violations} 

\noindent \parbox{3.8in}{ \openup\jot 
As illustrated in Fig.~4.1, in lowest order there are two diagrams for
($t$-channel) two-gluon exchange. The short-distance region $k^2_{\perp}
\sim Q^2 >> t$ gives non-leading twist behavior ($\int {d^2k_{\perp}
\over k^4_{\perp}} \sim 
{1 \over Q^2}$). Consequently leading-twist necessarily involves 
$k_{\perp} << Q^2$ for one gluon. 
If we consider 
$k_{\perp}^2 > \Lambda_{IR}~$ ($\equiv \Lambda_{QCD}~$ ?), 
the $ \Lambda_{IR}$ dependence cancels between 
the two diagrams. This cancellation is the well-known infra-red finiteness 
property of the BFKL pomeron }
 \parbox{2.2in} 
{\begin{center} 
\leavevmode
\epsfxsize=1.8in
\epsffile{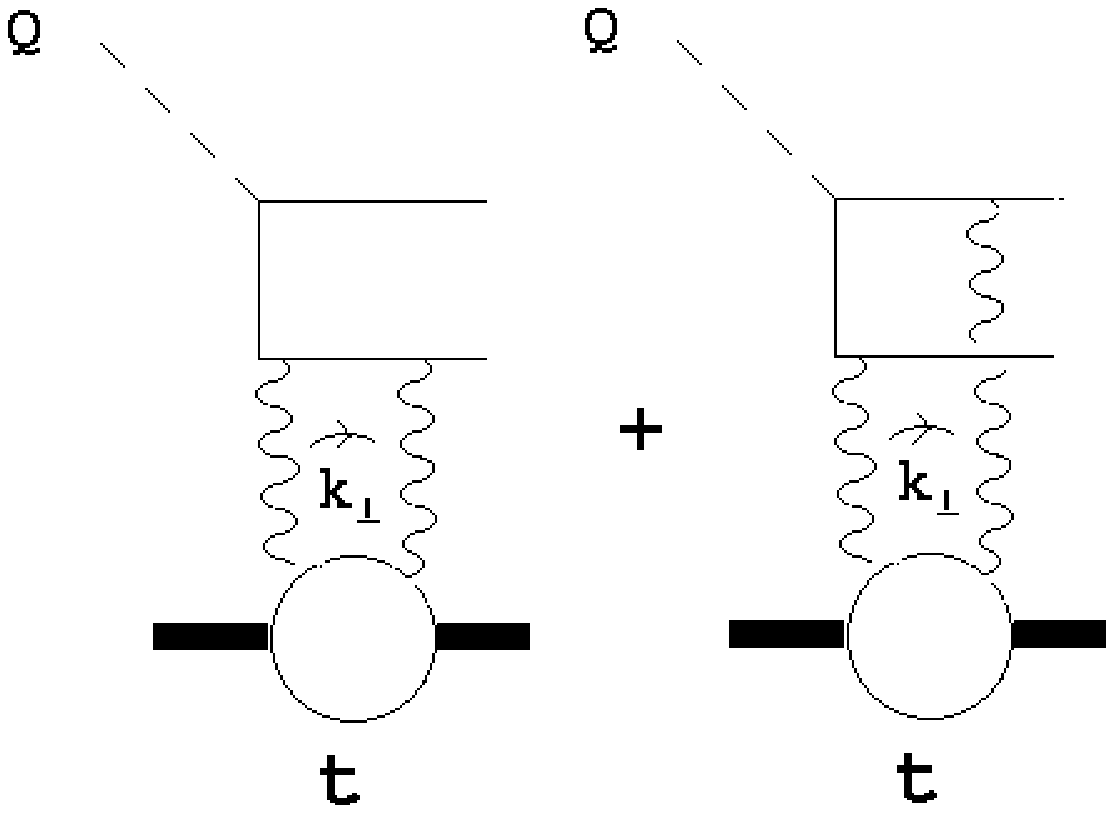}

Fig.~4.1 
\end{center}
}
(which follows
directly from gauge invariance).
The absence of $\Lambda_{IR}$ - dependence implies there is no 
scale for any potential $ln[Q^2]$ contributions, which indeed 
are absent~\cite{bw}. In effect, gauge invariance
implies that the $k_{\perp}$ regions for the two gluons can not be separated
sufficiently to produce a leading-twist, leading-order,
$\ln[Q^2]$. 

It is straightforward to extend the above argument to any number of 
perturbative gluons coupling directly to the quark loop and to
``soft Pomeron 
models'' defined via two (or more) ``non-perturbative'' gluon propagators. 
$ln[Q^2]$ dependence will arise, of course, when gluon exchanges between the 
gluons of Fig.~4.1 are taken into account, but this will be down by 
$O(\alpha_s)$
and will be too small to describe the scaling violations of Fig.~3.2.

As illustrated in Fig.~4.2, to obtain a large $\ln[Q^2]$ dependence in
``leading-oder'' and
\newline \parbox{4.1in} {\openup\jot 
 reproduce the $H1$ analysis, requires one hard gluon, 
with cut-off $\Lambda_{IR}$,
plus a  color compensating interaction with $\VEV{k_{\perp}}<<  \Lambda_{IR}$.
The above argument implies that gauge invariance will be violated  
unless the soft interaction is distinguished from single gluon 
exchange - via a quantum number. Even so, it is clearly non-trivial to
suppose that gauge invariance can be maintained when colored exchanges at
different scales are combined.} 
\parbox{1.9in}{ 
\begin{center} 
\leavevmode
\epsfxsize=1.6in
\epsffile{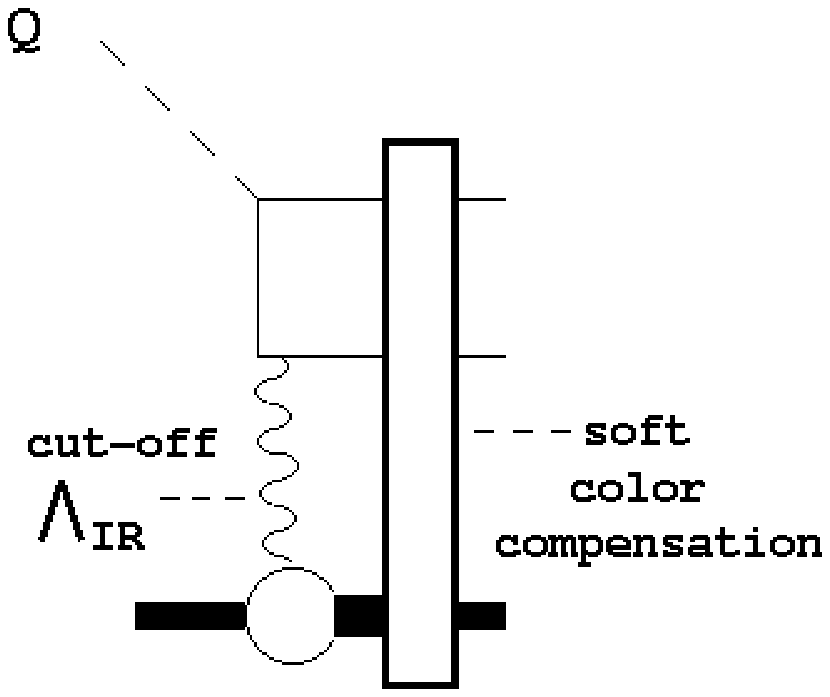}

Fig.~4.2 

\end{center}
}
\vspace{0.1in}

\section{Multi-Regge Theory - the Key Ingredients}

\noindent \parbox{3.8in}{ \openup\jot 	
{\it i)  Angular Variables } - For the N-point amplitude we write 
\newline \centerline{$ M_N(P_1,..,P_N) \equiv
M_N\left(t_1,..,t_{N-3},g_1,..,g_{N-3}\right)$}
where, in the notation of Fig.~5.1, 
$t_j=Q_j^2$ and $g_j \in $ the little group of $Q_j$, i.e. 
$g_j \in$ SO(3) or 
$g_j \in$ SO(2,1) for $t_j 
{\raisebox{1mm}{\centerunder{$\scriptscriptstyle 
>$}{$\scriptscriptstyle <$}}}~ 0.$ 
There are 
N-3 $~t_i$ variables, 
N-3 $~z_j ~(\equiv \cos\theta_j)$ variables 
and 
N-4 $~u_{jk}~ (\equiv e^{i(\mu_j - \nu_k)})$ variables.
}
\parbox{2.2in}{
\begin{center}
\leavevmode
\epsfxsize=1.8in
\epsffile{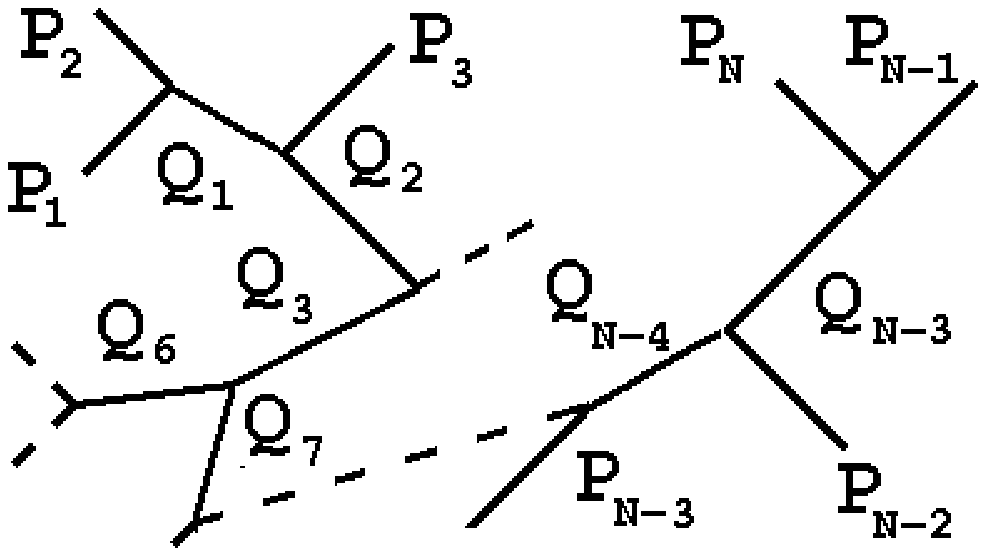}

Fig.~5.1
\end{center}
}

\noindent ii) {\it A Multi-Regge Limit } is defined by  
$z_j \to \infty ~\forall j$. In {\it a Helicity-Pole Limit}, some $u_{jk} 
\to \infty$ and some $z_j \to \infty$.

\noindent iii) {\it Partial-wave Expansions}. Using 
$ f(g)=\sum^\infty_{J=0}\,\sum_{|n|,|n'|<J}D^J_{nn'}(g)a_{J nn'}$, 
for a function $f(g)$ defined 
on SO(3), leads to 
$M_N(\til{t},\til{g})=\sum_{\til{J},\til{n},\til{n'}}
\prod_i~D^{J_i}_{n_in_i'}(g_i)~ a_{\til{J},\til{n},\til{n'}} (\til{t})
$

\noindent iv) {\it Asymptotic Dispersion Relations}. 
We can write $~~~~M_N~=~\sum_{C}M_N^C +M^0~~~~
$ where
\newline \centerline{$  M_N^C ~=~ {1\over (2\pi i)^{N-3}}
\int \frac{ ds'_1\ldots ds'_{N-3}\Delta^C(
..t_i.,..u_{jk}.,..s'_i.)}
{(s'_1-s_1)(s'_2-s_2)\ldots (s'_{N-3}-s_{N-3})} $}
\newline and $\sum_{C}$ is over all sets of (N-3) Regge limit asymptotic cuts.
($M^0$ is non-leading in the multi-regge limit.) The resulting separation into
(hexagraph) spectral components is crucial for the development of
multiparticle complex angular theory. 

\noindent v) {\it Sommerfeld-Watson Representations 
of Spectral Components} e.g.
\vspace{0.1in}
\newline \centerline{$M^C_4={1\over 8}\sum_{{\scriptstyle N_1, N_2}} \int
{dn_2  dn_1 dJ_1 ~u_2^{n_2} u_1^{n_1}
d^{J_1}_{0,n_1}(z_1)
d^{n_1+N_2}_{n_1,n_2}(z_2)d^{n_2+N_3}_{n_2,0}(z_3)
\over
\sin\pi n_2\sin\pi(n_1-n_2)\sin\pi(J_1-n_1)}~a^C_{N_2N_3}(J_1,n_1,n_2,
\til{t})
$}
\newline from which the form of multi-Regge 
behaviour in any limit can be extracted. 

\noindent vi) {\it $t$-channel Unitarity in the $J$-plane}
~~~Multiparticle unitarity in every $t$-channel can be partial-wave
projected, diagonalized, and continued to complex $J$ in the form 
\vspace{0.1in}
\newline \centerline {$ a^+_J - a^-_J= i\int d\rho \sum_{\til{N}} 
\int {dn_1 dn_2  \over 
sin\pi(J-n_1-n_2) }\int {dn_3 dn_4 \over sin\pi(n_1 -n_3 -n_4)} ~\cdots 
~a^+_{J\til{N}
\til{n}}a^-_{J\til{N}\til{n}} 
$}
\newline Regge poles at $n_i=\alpha_i$, together with the phase-space 
$\int d\rho $ and the ``nonsense poles'' at 
$J= n_1 +n_2 -1, n_1=n_3 + n_4 -1 , ~...~$ generate Regge cuts. 

\noindent vii) {\it Reggeon Unitarity }.
In ANY partial-wave amplitude, the $J$-plane discontinuity
due to $M$ Regge poles $\til{\alpha}= (\alpha_1, \alpha_2, \cdots \alpha_M)$
is given by the reggeon unitarity equation 
\newline \centerline{ $ 
\centerunder{disc}{\raisebox{-3mm}{$\scriptstyle J=\alpha_M(t)$}}~~ 
a_{\til{N} \til{n}}(J) 
~=~ {\xi}_{M} \int d\hat\rho~
a_{\til{\alpha}}(J^+)
a_{\til{\alpha}}(J^-)
{\delta\left(J-1-\sum^M_{k=1}
(\alpha_k-1)\right)\over \sin{\pi\over 2}(\alpha_1-\tau'_1)\ldots\sin{\pi\over
2}(\alpha_M-\tau'_M) }
$ }
\newline Writing $t_i=k_i^2~~$ (with 
$\int dt_1 dt_2 \lambda^{-1/2}(t,t_1,t_2) =2\int d^2 k_1 d^2 k_2 \delta^2(k 
- k_1 - k_2) $), $\int d \hat{\rho} $ can be written in terms of
two dimensional ``$~k_{\perp}$'' integrations, 
anticipating the results of 
direct ($s$-channel) high-energy calculations (leading-log, next-to-leading 
log etc.). 
Because the gluon ``reggeizes'' in perturbation theory, i.e. 
is a Regge pole in the $J$-plane, reggeon unitarity is a 
powerful constraint on the higher-order contributions of multigluon
exchange. 
\vspace{0.1in}

\section{Reggeon Diagrams for Multi-Regge Limits}
 
Reggeon unitarity is particularly powerful when applied to multiparticle
``helicity-pole'' limits. If the SW representation shows 
only one partial-wave amplitude is involved, reggeon unitarity
implies that the limit is fully described by two-dimensional ``$~k_{\perp}~$'' 
diagrams. (In fact the ``physical'' transverse planes
actually contains lightlike momenta.) 

\noindent E.g. for an 8-point amplitude introduce angular 
variables as in Fig.~6.1 and consider 
\newline \parbox{3.8in}{ \openup\jot 
the ``helicity-flip'' helicity-pole limit in which 
\newline $ z,u_1,u_2,u_3,u_4~\to \infty~$. The behavior of invariants is 
$$
\eqalign{&P_1.P_2 \sim u_1u_2, ~~P_1.P_3 \sim u_1zu_3, ~~P_2.P_4 \sim u_2u_4,
\cr
&P_1.Q_3 \sim u_1z, ~~Q_1.Q_3 \sim z, ~~P_4.Q_1 \sim zu_4 
~ ~~ \cdots \cr
&~ P_1.Q,~P_2.Q,~ P_3.Q,~P_4.Q  ~~\hbox{{\it finite }} }
$$}
\parbox{2.2in}{
\begin{center}
\leavevmode
\epsfxsize=1.8in
\epsffile{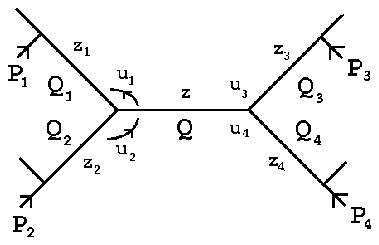}
\newline \centerline{Fig.~6.1}
\end{center}}
\newline 
Since the S-W representation shows that only 
one partial-wave amplitude is involved, 
a simple reggeon unitarity equation holds in all channels. As a result 
the complete multi-Regge behavior can be described by $k_{\perp}$
integrals of the form illustrated in Fig.~6.2, where
$~~{\raisebox{-1mm}{\epsfxsize=0.25in \epsffile{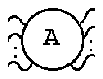}}}$
\newline
\parbox{2.2in}{
\begin{center}
\leavevmode
\epsfxsize=1.8in
\epsffile{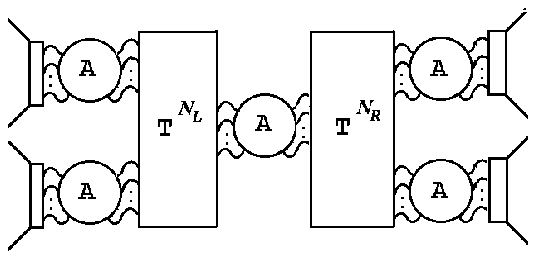}
\newline {Fig.~6.2} \end{center}
}
\parbox{3.8in}{ \openup\jot 
contains all elastic scattering reggeon diagrams. 
$T^L,T^R$ contain 
connected and disconnected interactions that 
involve both 
elastic scattering (s-channel ``helicity non-flip'') 
reggeon vertices and also new 
``helicity-flip'' vertices. }

Helicity-flip vertices play a crucial dynamical role. They do 
not appear in either elastic scattering 
or multi-Regge production processes. Such vertices are most simply 
isolated in 
\newline \parbox{4.8in}{ \openup\jot a ``non-planar''
triple-regge limit involving three light-cone momenta. In 
the notation of
Fig.~6.3 we can define a non-planar triple-regge limit by 
\vspace{0.05in}
\newline \centerline{$
P_1\to (p_1,p_1,0,0) ~~~p_1 \to \infty , ~~~~~~ Q_1\to (0,0,q_2, -q_3)$}
\newline \centerline{$P_2\to (p_2,0,p_2,0)~~~p_2 \to \infty ,
~~~~~~ Q_2\to (0,-q_1,0,q_3)$}
\newline \centerline{$P_3\to (p_3,0,0,p_3)~~~p_3 \to \infty , 
~~~~~~ Q_3\to  (0,q_1,-q_2,0)$}
}
\parbox{1.2in}{
\begin{center}
\leavevmode
\epsfxsize=0.9in
\epsffile{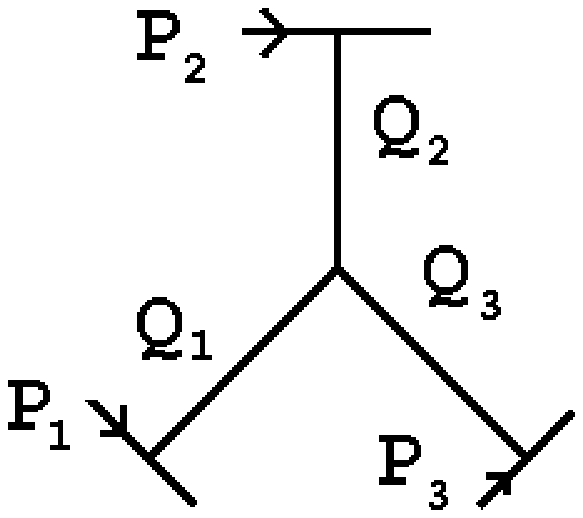}
\newline Fig.~6.3
\end{center}
}

\section{Reggeized Gluon Helicity-Flip Vertices Involving Massless Quarks} 

\noindent \parbox{4.3in}{\openup\jot Consider three quarks scattering via 
gluon exchange with a quark loop triple-gluon coupling as illustrated in
Fig.~7.1. The non-planar
triple-regge limit discussed above gives {\small
$~ g^6 { p_1p_2p_3 \over t_1 t_2 t_3 } ~\Gamma_{1^+2^+3^+}(q_1,q_2,q_3)
$ }
where $t_i = Q_i^2, ~ \gamma_{i^+} = \gamma_0 + \gamma_i $ and
$\Gamma_{\mu_1 \mu_2 \mu_3}$ is given by the quark triangle diagram, i.e.
\vspace{0.1in}
{\small \newline \centerline{$
\Gamma_{\mu_1 \mu_2 \mu_3} = i\int {  d^4 k~ Tr \{ \gamma_{\mu_1}
(\st{q}_3 + \st{k} + m ) \gamma_{\mu_2} (\st{q_1} + \st{k} + m ) 
\gamma_{\mu_3} (\st{q}_2 + \st{k} + m) \} 
\over [ (q_1 + k)^2 - m^2 ][ (q_2 + k)^2 - m^2 ]
[ (q_3 + k)^2 - m^2 ]}
$} } }
\parbox{1.7in}{
\begin{center}
\leavevmode
\epsfxsize=1.4in
\epsffile{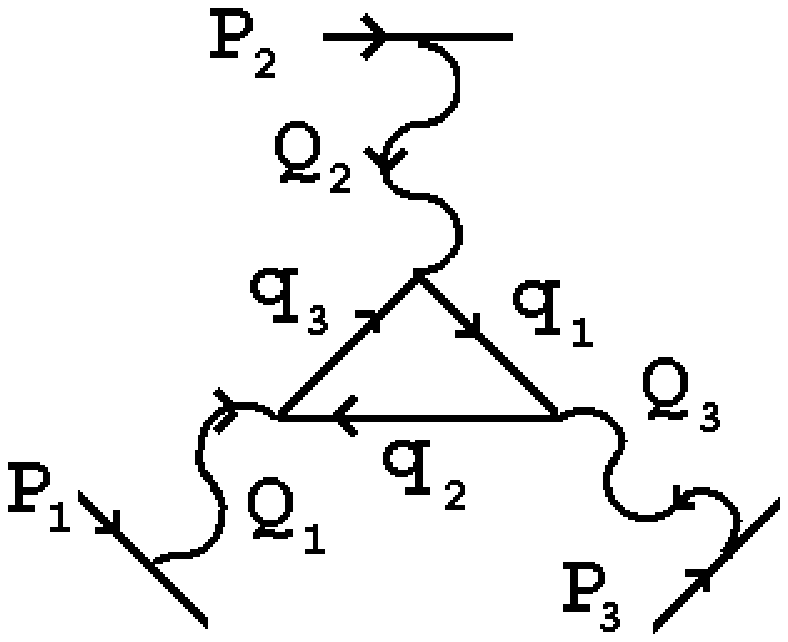}
\newline $~$
\newline \centerline{Fig.~7.1}
\end{center}
}
It is a crucial property of the ``$O(m^2)$'' part of 
$~\Gamma_{1^+2^+3^+}~$ that the limits $q_1, q_2, q_3 \sim Q
\to 0$, $m \to 0$ do not commute. (An ``infra-red 
anomaly'' due 
to the triangle Landau singularity).  
\vspace{0.1in}
\centerline{$
\Gamma_{1^+2^+3^+,m^2} {\centerunder{$\sim$}{\raisebox{-4mm} 
{$Q \sim 0$} }}~Q ~i~m^2 \int {d^4k \over [ k^2 - m^2 ]^3 }
 ~~~{\centerunder{$\longrightarrow $}{\raisebox{-4mm} {$ m \to 0 $}} }~
  R ~Q ~\neq~~0
$} 

After adding color factors and summing diagrams, $\Gamma_{1^+2^+3^+,m^2}$ 
survives only in triple-regge vertices that couple reggeon states which all 
have ``anomalous color parity'', i.e.
\newline \parbox{4.3in}{ \openup\jot 
$C_c= - \tau$. ($\tau =$ signature and color parity is defined via
$A^i_{ab} \to - A^i_{ba}$ for gluon color matrices.) An important example is  
shown in Fig.~7.2. In this case 
each three reggeon state has odd signature but even color parity, 
e.g. $f_{ijk}d_{jrs}A^kA^rA^s$ (c.f. the winding-number current
$K^i_{\mu}= \epsilon_{\mu \nu \gamma 
\delta}f_{ijk}d_{jrs}A^k_{\nu}A^r_{\gamma}A^s_{\delta}$).
In these circumstances, the survival of 
$O(m^2)$ processes as
$m \to 0$ reproduces the chirality violation of
instanton interactions.}
\parbox{1.7in}{
\begin{center}
\leavevmode
\epsfxsize=1.4in
\epsffile{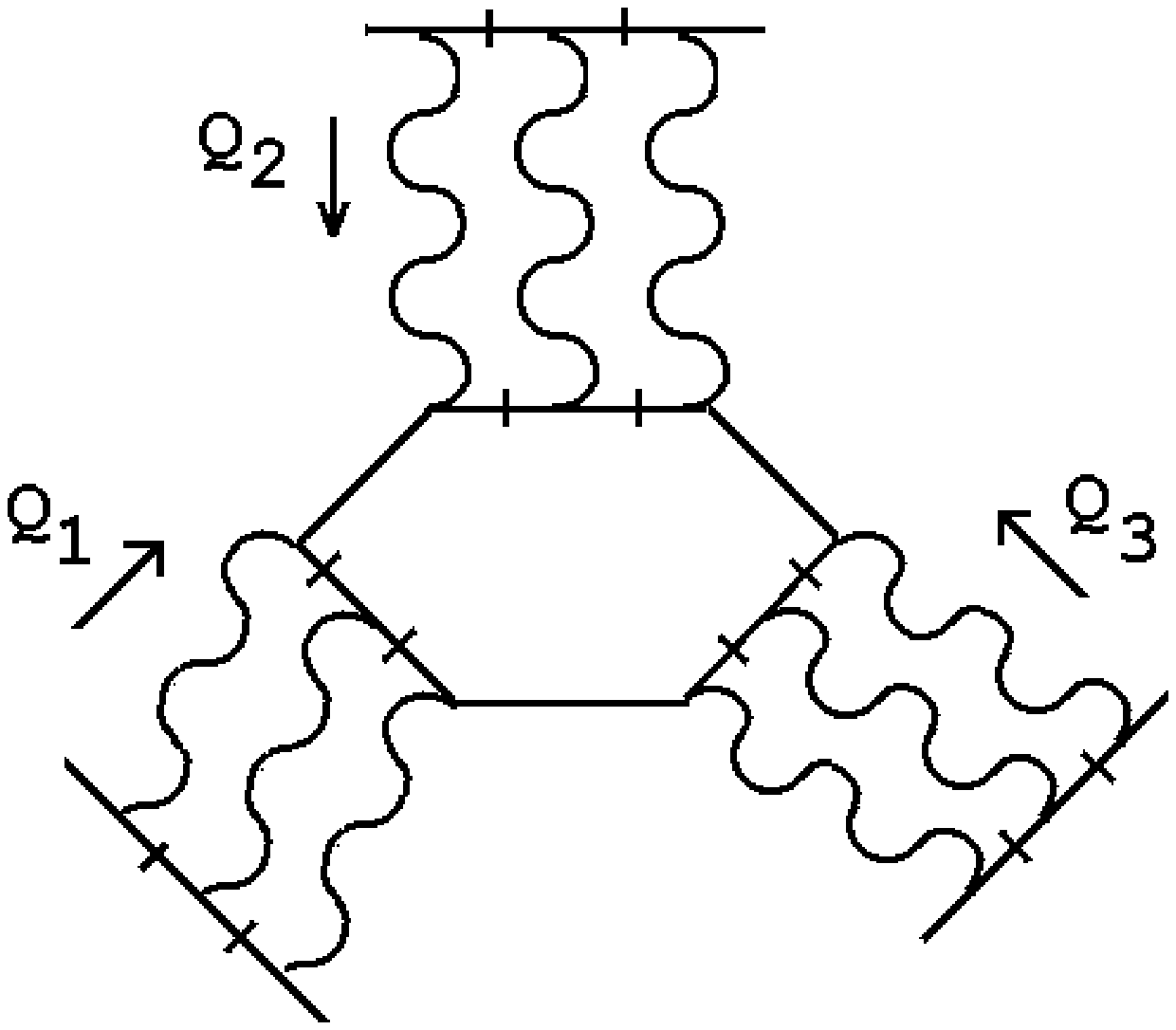}
\newline Fig.~7.2
\end{center}
}

Note that $\Gamma_{1^+2^+3^+,m^2}$ has $[k_{\perp}]$-dimension 1 !!
Normal reggeon interactions have $[k_{\perp}]$-dimension - 2 .
When combined with transverse momentum conservation i.e.
$\delta^2(k_{\perp})$, the normal interactions naturally produce a
scale-invariant (and even conformal invariant) 
\newline \parbox{4.7in}{ \openup\jot 
massless theory. Because of it's anomalous dimension, 
$\Gamma_{1^+2^+3^+,m^2}$ can only appear in special multi-Regge limits where 
the large invariants contribute an 
an extra momentum 
factor. The non-planar 
triple-regge limit discussed above is an example. In the notation of 
Fig.~7.3 {\small
\vspace{0.1in}
\newline $ p_1p_2p_3 Q_i \leftrightarrow~
 (p_1p_3)^{1/2}(p_2p_3)^{1/2} (p_1p_2)^{1/2} Q_i 
\equiv (s_{31})^{1/2}( s_{23})^{1/2}(s_{12})^{1/2} Q_i $
}
}
\parbox{1.3in}{
\begin{center}
\leavevmode
\epsfxsize=1in
\epsffile{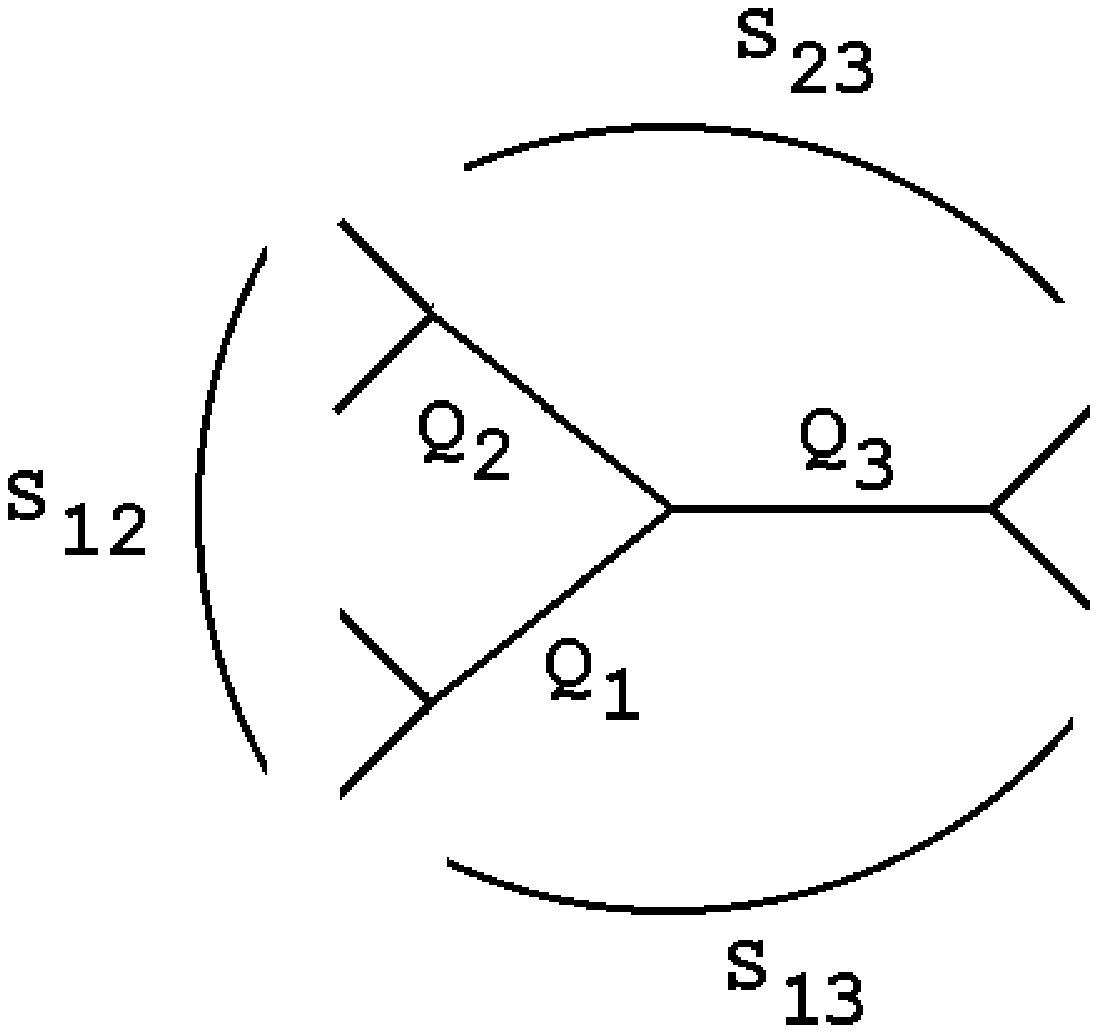}
\newline Fig.~7.3
\end{center}
}

\section{Infra-red Divergences}

The anomalous dimension $\Gamma_{1^+2^+3^+,m^2}$ 
also implies a ``violation'' of 
conventional reggeon Ward identities, i.e. when $m \to 0$, the vertex does 
not vanish sufficiently fast when all $Q_i \to 0$.
(When $m \neq 0$, other
diagrams combine to produce the normal reggeon Ward identities.)
As a consequence, when the gluon mass is zero, an infra-red divergence
appears as $m \to 0$ in a large class of non-planar multi-regge diagrams
where $~Q_1 \sim Q_2 \sim Q_3 \sim 0~$ is 
\newline 
\parbox{2.2in}{ 
\leavevmode
\epsfxsize=1.8in 
\epsffile{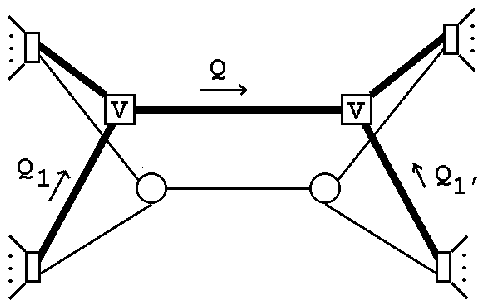}
\newline \centerline{Fig.~8.1 $~~~~~~$ }
}
\parbox{3.8in}{\openup\jot 
part of the integration region. A candidate diagram 
is shown in Fig.~8.1.
Both \epsfxsize=0.2in \epsffile{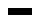}$~$ and
\epsfxsize=0.2in \epsffile{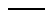}$~$ are reggeon states.
A divergence will occur for 
$Q,Q_1,Q_{1'} \sim 0$ as $m \to 0$ if the vertex $V$ contains 
$\Gamma_{1^+2^+3^+,m^2}~$. This requires that 
\epsfxsize=0.2in \epsffile{dss15.ps} ~be an anomalous
color parity 
reggeon state.
To discuss whether the divergence is cancelled by other diagrams 
requires a systematic analysis which begins with the SU(3) }
\newline gauge 
symmetry initially restored only to SU(2) (c.f. an instanton is
associated with an SU(2) subgroup). 
In this case the divergence occurs when 

\noindent \epsfxsize=0.2in \epsffile{dss16.ps} ~~is an SU(2) singlet state 
containing 
one or more massive reggeized gluons (or quarks) 

\noindent \epsfxsize=0.2in \epsffile{dss15.ps} ~is an SU(2) singlet combination 
of massless gluons with 
$~C_c= -\tau = +1~ \equiv~$ (``anomalous odderon'').

\noindent \parbox{2.9in}{
\leavevmode
\epsfxsize=2.5in
\epsffile{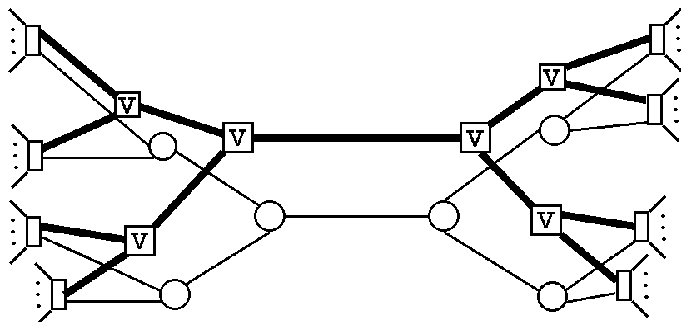}
\newline \centerline{Fig.~8.2 $~~~$}
}
\parbox{3.1in}{ \openup\jot 
Higher-order diagrams containing a divergence include those of the form 
shown in Fig.~8.2. There is always an overall 
logarithmic divergence from that part
of the integration region 
where all the $Q_i$ entering each $V$ 
vertex vanish. 
``Physical amplitudes'' are obtained by extracting the coefficient of the
divergence. In this 
}
coefficient all anomalous odderon reggeon states carry 
zero $k_{\perp}$ and so, effectively, the definition of reggeon states 
includes ``an anomalous odderon reggeon condensate''.
There is also, as illustrated in Fig.~8.3, an SU(2) singlet ``parton
process'' which carries the kinematic properties of the dynamical
interaction.  Since the condensate background results from the 
quark triangle anomaly it (qualitatively) can be thought of as originating 
from instanton interactions. 

\noindent 
\parbox{2.9in}{ 
\leavevmode
\epsfxsize=2.5in
\epsffile{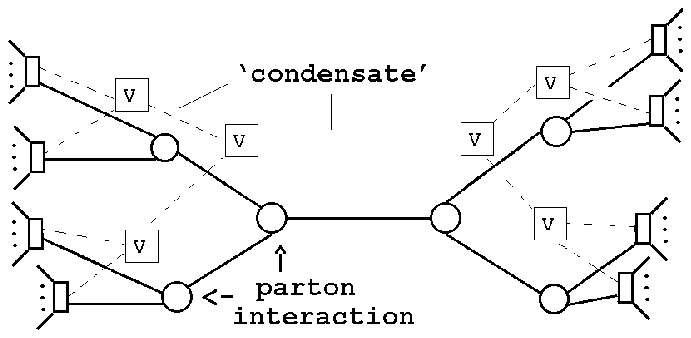}
\newline \centerline{ Fig.~8.3 }
}
\parbox{3.1in}{
The ``pomeron'' appears as an SU(2) singlet
reggeized gluon (with $C_c = \tau = - 1$) in the $C_c = - \tau = +1$
condensate, i.e. an even signature Regge pole with $C_c = - \tau = - 1$.
``Hadrons'' are similarly (constituent) quark reggeon states in the
condensate background. (It appears that ``stability'' within the condensate
requires chiral symmetry breaking, but I will not discuss } 
this here. Note, however, that $C_c= -1$ for the pomeron implies that hadrons
can not be eigenstates of 
$C_c$. This does not imply there is no charge conjugation symmetry.)
All features~\cite{arw2} of super-critical pomeron RFT are present, 
but I will not discuss the details. 

Restoration of SU(3) gauge symmetry (i.e. the decoupling of a ``Higgs'' scalar) 
is straightforward provided there is a cut-off
$|k_{\perp}| < \Lambda_{\perp}$. The most important point is that the
pomeron is described by RFT and carries a crucial remnant of the
construction i.e. odd SU(3) color charge parity. An immediate implication
is that the BFKL Pomeron does not appear (since it has $C_c = +1$).
To remove $\Lambda_{\perp}$ the pomeron should be
Critical. (This is related to the quark content of the 
theory and again we will not discuss it here.) A-priori, at a fixed value of 
$\Lambda_{\perp}$, the theory may be above, or below, the critical point.

\section{Deep-Inelastic Scattering }

For this discussion we consider the SU(3) gauge symmetry to be restored and 
assume the critical cut-off $\Lambda_{\perp c}$ is above the physical cut-off
so that the full theory is sub-critical. DIS diffractive 
scattering will expose the simplest perturbative contribution to the 
pomeron. $C_c = -1$ determines this to be 
four gluon exchange. A 
``non-perturbative'' coupling $I_c$ to chirality-violating processes must be 
involved. We assume the chirality-violating scale is smaller than the 
``perturbative'' $Q^2$ scale. 

\noindent \parbox{3.5in}{ \openup\jot $~~~$ As illustrated in Fi.~9.1, there
are two distinct reggeon/gluon diagram contributions, distinguished by 
whether the anomalous color parity state coupling via $I_c$ involves 
three or two reggeized gluons.
In the Regge limit (finite $Q^2$) the $8_s$ 
two gluon state gives an
even signature~\cite{bw} 
reggeon with a ``pointlike'' 
coupling to the quark loop. 
In higher-orders, the $8_A$ single
}
\parbox{2.5in} { \begin{center} 
\leavevmode
\epsfxsize=1.9in
\epsffile{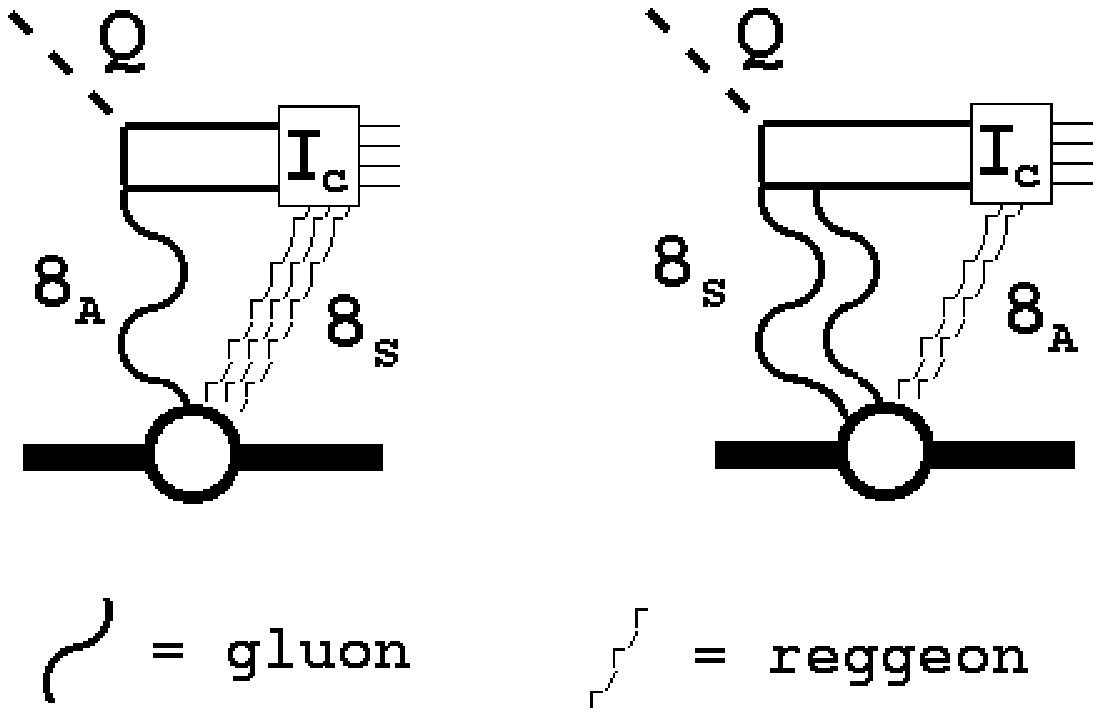}
\newline Fig.~9.1
\end{center}
}
\newline 
gluon and the $8_s$ two 
gluon configurations reggeize with identical trajectories. Hence 
regge-limit infra-red divergences (due to the anomalous reggeon
states) from the two contributions can directly cancel. 

At large $Q^2$ the first diagram, with the single hard gluon, dominates. The 
same effect is achieved by breaking the 
SU(3) gauge symmetry to SU(2). This is why we can say that the pomeron is
``in the super-critical phase'' at the deep-inelastic scale. The result is
the hard gluon behavior seen by H1 !

\section*{Acknowledgements}

I am particularly grateful to Mark W\"usthoff for extensive discussions.

\end{document}